\documentclass[twocolumn,prb,reprint,superscriptaddress,floatfix,amsmath,amssymb,aps]{revtex4-2}
\pdfoutput=1

\usepackage[utf8]{inputenc}
\usepackage[caption=false,subrefformat=parens,labelformat=parens]{subfig}
\usepackage{graphicx}
\usepackage{float}
\usepackage{amsmath}
\usepackage{dcolumn}
\usepackage{bm}
\usepackage{xcolor}
\usepackage{xfrac}
\usepackage[colorlinks=true, allcolors=blue]{hyperref}
\usepackage{chemformula}
\usepackage{siunitx}

\newcommand{\svs}{\ch{ScV6Sn6}}
\newcommand{\svcs}{Sc(V$_{1-x}$Cr$_x$)$_6$\ch{Sn6}}

\begin{document}
\title{Multiple closely spaced transitions and multi-band Hall response in clean \svs}

\author{Jonathan M. DeStefano}
\affiliation{Department of Physics, University of Washington, Seattle WA, 98112, USA}
\author{Elliott Rosenberg}
\affiliation{Department of Physics, University of Washington, Seattle WA, 98112, USA}
\author{Chaowei Hu}
\affiliation{Department of Physics, University of Washington, Seattle WA, 98112, USA}
\author{Xiaodong Xu}
\affiliation{Department of Physics, University of Washington, Seattle WA, 98112, USA}
\affiliation{Department of Material Science and Engineering, University of Washington, Seattle WA, 98112, USA}
\author{Jiun-Haw Chu}
\email{jhchu@uw.edu}
\affiliation{Department of Physics, University of Washington, Seattle WA, 98112, USA}

\date{\today}

% Elliott will look for ACER data

\begin{abstract}

The kagome metal \svs\ has attracted attention as a platform for exploring the interplay between charge density wave (CDW) order and symmetry-breaking phenomena, including a recently reported intermediate phase and a low-field Hall anomaly that has been attributed to an anomalous Hall effect (AHE). The interpretation of both observations has been limited by the modest sample quality achieved by previous growth procedures, which produced crystals with in-plane residual resistivity ratios (RRR) of at most $\approx$9. Here, we report a simple modification of the flux growth procedure that yields \svs\ single crystals with RRR exceeding 50, more than five times the previous highest reported value, and use this expanded mobility range to revisit both the symmetry and the magnetotransport of the CDW phase. We resolve a sequence of closely spaced transitions in the immediate vicinity of $T_{CDW}$ that emerges above a sharp threshold of RRR $\approx 4$, and demonstrate through elastoresistivity that the intermediate phase breaks the three-fold rotational symmetry of the parent lattice. We examine the Hall response from both the parent samples across the full RRR range as well as Cr-doped samples, and conclude it is quantitatively inconsistent with an intrinsic AHE and is instead explained by ordinary multi-band transport involving small, high-mobility pockets identified through quantum oscillations. These results refine the symmetry-breaking landscape of \svs\ and establish systematic mobility tuning as a diagnostic for disentangling an intrinsic AHE from multi-band Hall contributions in kagome CDW systems.

\end{abstract}

\maketitle

\section{Introduction}

Materials hosting kagome lattices have emerged as a fertile ground for studying the interplay between topology, geometric frustration and electronic correlations. Among them, kagome metals hosting charge density waves (CDWs) have attracted particular attention, owing to their rich phase diagrams and the variety of intertwined symmetry-breaking phases that emerge below the CDW transition~\cite{Ortiz2020,Jiangetal2021,Teng2022,svs_discovery}. A notable example is the \textit{A}\ch{V3Sb5} (\textit{A} = K, Rb, Cs) family~\cite{WilsonOrtiz2024}, in which the CDW coexists with a low temperature superconducting state ($T_c <$ \SI{3}{K}) and has been proposed to break both rotational symmetry and time-reversal symmetry (TRS)~\cite{Zhao2021,MuSR2022,CVSAMRO2021,CVSChiral2022,LiangWu2022}, motivating a host of exotic scenarios including orbital loop currents, pair density wave and electronic nematicity. Whether TRS and rotational symmetry are genuinely broken in the CDW phase, however, remains contested, with different probes yielding conflicting conclusions\cite{Saykin2023,Farhangetal2023,kapitulnik_SVS,CVS_nonematicity,ColossalC-axisCVS}. A central piece of evidence cited in support of TRS breaking is a low-field Hall anomaly that cannot be described by a simple two-band model, which has been interpreted as a field-induced anomalous Hall effect (AHE)~\cite{CVSAHE, KVSAHE}. However, it was demonstrated that this anomaly instead originates from a set of small (quantum oscillation frequency $<$ \SI{100}{T}) Fermi pockets with high mobilities~\cite{CVS_tinypockets}. This conclusion was reached through magnetotransport measurements on electron irradiated samples of \ch{CsV3Sb5}, in which the carrier mobility was systematically reduced (correspondingly reducing the RRR) without altering the electronic structure. This result highlights the importance of varying the degree of disorder as a means of disentangling the microscopic origin of magnetotransport anomalies in multi-band kagome systems.

% should we mention the high temperature fluctuations in SVS? Implies phases with very similar free energies

Like the \textit{A}\ch{V3Sb5} compounds, \svs\ hosts kagome layers comprised solely of V ions and undergoes a CDW transition below $\approx$\SI{90}{K}\cite{svs_discovery}. Several key differences nevertheless distinguish the two systems. No superconductivity has been observed in \svs\ down to the lowest measure temperatures nor under pressure~\cite{svs_pressure}, and the CDW adopts a $\sqrt{3} \times \sqrt{3} \times 3$ ordering driven primarily by structural frustration\cite{rattlingchain,korshunov_softening_2023,tan_abundant_2023,cao_competing_2023}, in contrast to the in-plane $2 \times 2$ ordering of the \textit{A}\ch{V3Sb5} family, which nests the van Hove singularity at the Brillouin zone boundary~\cite{cvs_fermisurfacenesting}. Notably, the electronic structure of \svs\ closely resembles that of the \textit{A}\ch{V3Sb5} compounds, so a van Hove nesting mechanism would be expected to select the same in-plane wave vector; the distinct ordering vector of \svs\ therefore points to a lattice-driven instability rather than a purely electronic one~\cite{ huPhononPromotedCharge2024}. Despite these distinctions, signatures of TRS breaking have likewise been reported in \svs\ on the basis of muon spin relaxation measurements~\cite{hiddenmagnetism_svs} and a low-field Hall anomaly~\cite{QuantumOscillationsRevealing_SVS}, suggesting a kagome-derived mechanism that transcends the specific CDW wave vector. More recently, a second phase transition has been identified at a temperature $T^*$ a few K below $T_{CDW}$~\cite{intermediatenematic_SVS}. The intermediate phase between $T^*$ and $T_{CDW}$ displays a pronounced spontaneous in-plane anisotropy and was observed most prominently in a sample with a RRR of 9, which is the highest value reported in the literature to date\cite{svs_discovery,universalsublinearresistivity_SVS_CVS,SVS_PG,QuantumOscillationsRevealing_SVS}. This raises the question of whether the intermediate phase has been suppressed by disorder in lower-quality samples. An investigation of \svs\ across a wide range of RRR therefore serves a twofold purpose: it tests the disorder sensitivity of the intermediate nematic phase, and it sheds light on the origin of the low-field Hall anomaly.

In this paper, we present a simple synthesis method that tunes the RRR of \svs\ over an unprecedented range, from less than 3 to greater than 50, more than five times the highest value previously reported~\cite{intermediatenematic_SVS}. Electrical transport measurements across this series reveal a sharp threshold at RRR $\approx$ 4, separating samples that exhibit a single transition at $T_{CDW}$ from those that exhibit multiple closely spaced transitions. Remarkably, many samples with RRR $>$ 20 display more than two transitions, pointing to an unusually rich energy landscape in the vicinity of $T_{CDW}$. Symmetry-resolved elastoresistivity shows a dramatic enhancement in the intermediate phase, confirming its broken rotational symmetry nature. To complement the disorder-tuning approach, we also grew single crystals of \svcs, in which Cr substitution nominally electron-dopes \svs\ while simultaneously reducing the mobility through substitutional disorder. Systematic magnetotransport and quantum oscillation measurements on both \svs\ and \svcs\ reveal small, light pockets, in good agreement with previous density functional theory (DFT) calculations. The Hall effect in \svs\ deviates from a standard two-band description, particularly in high RRR samples, as expected for a multi-band system with strongly disparate mobilities. Furthermore, Hall measurements of \svcs\ reveal that even at \SI{14}{T} the system has not reached the high-field regime, and thus previous extractions of an ``AHE" in \svs\ are not well supported. In total, our results indicate that the CDW in \svs\ exists in an unusually rich energy landscape with several nearby phases, with no clear evidence of TRS breaking in our magnetotransport data.

\section{Results}

\subsection{Phase transitions}
\label{sub:transitions}

\begin{figure*}
    \centering
    \includegraphics[width=0.9\textwidth]{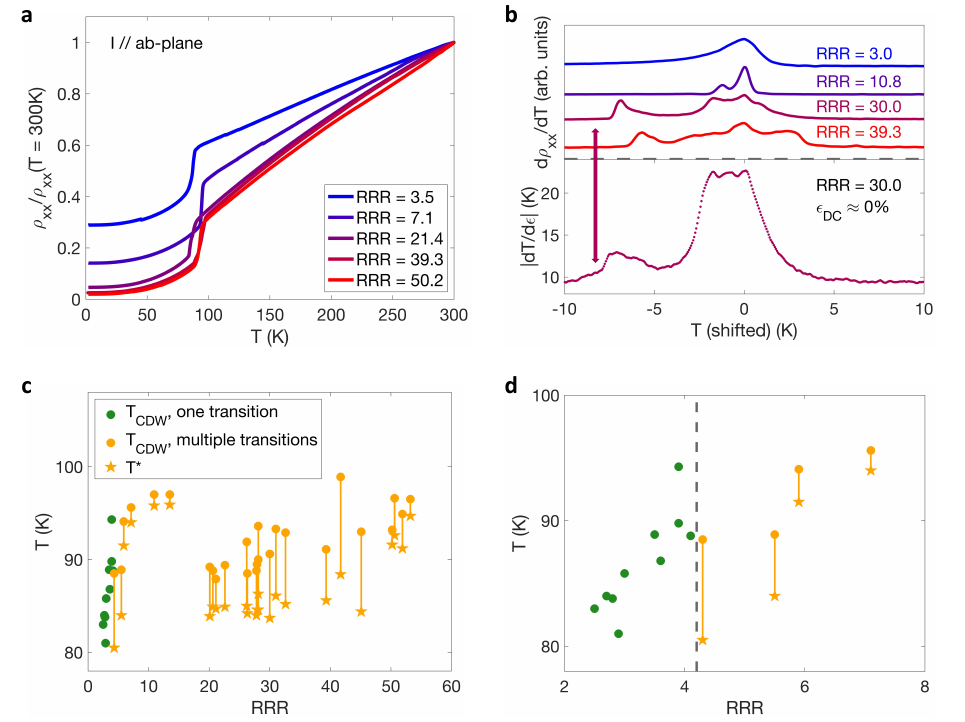}
    \caption{\textbf{Phase transitions in clean \svs.} \textbf{a,} In-plane resistivity $\rho_{xx}$, normalized by its \SI{300}{K} value, as a function of temperature for \svs\ samples with RRR ranging from 3.5 to 50.2. \textbf{b,} Top: $\frac{d\rho_{xx}}{dT}$ near $T_{CDW}$ for several samples, offset vertically for clarity and shifted in temperature so that the high-temperature peak aligns with \SI{0}{K}.  A single peak in the RRR~=~3.0 sample splits into progressively more features with increasing RRR. Bottom: elastocaloric effect ($|dT/d\epsilon|$) for RRR $= 30.0$ sample on the same shifted temperature scale, measured under near zero DC strain ($\epsilon_\text{DC} \approx 0$\%). Two clusters of peaks are resolved, in agreement with $d\rho_{xx}/dT$ and confirming the thermodynamic origin of the splitting. \textbf{c,} Transition temperatures versus in-plane RRR for nearly 40 samples of \svs. Green points denote samples with a single transition; orange points denote samples with multiple transitions with dots and stars indicating the onset and offset of the transition range identified in $\frac{d\rho_{xx}}{dT}$. \textbf{d,} A zoom in on the low RRR regime, where a vertical dashed line marks the sharp boundary RRR~$\approx 4$ separating single- and multi-transition samples.}
    \label{fig:RRR_phasetransitions}
\end{figure*}

Samples of \svs\ with varying degree of disorder were grown by changing the percentage of \ch{H2}-cleaned V pieces included in the starting materials (see Sec.~\ref{sec:methods} for the full experimental details). The relative degree of disorder is inferred from the RRR. Fig.~\ref{fig:RRR_phasetransitions}a presents the temperature dependence of in-plane resistivity ($\rho_{xx}$) normalized by the \SI{300}{K} value, for samples spanning a wide range of RRR (defined as RRR $=\frac{\rho(T=300K)}{\rho(T=3.5K)}$, with the low temperature value chosen to avoid the influence of superconducting Sn flux inclusions below \SI{3.5}{K}). The CDW transition is clearly resolved as a sudden drop in resistivity upon cooling through $T_{CDW}$ $\approx$\SI{90}{K}. Notably, the size of the drop in $\rho_{xx}$ at $T_{CDW}$ does not change significantly with RRR; instead, the slope of $\rho_{xx}$ above $T_{CDW}$ increases significantly as a function of RRR. 
%This could be related to the high temperature CDW fluctuations that have previously been shown to impact the magnetotransport of \svs~\cite{SVS_PG}. 

The top panel of Fig.~\ref{fig:RRR_phasetransitions}b displays $\frac{d\rho_{xx}}{dT}$ near $T_{CDW}$ for several samples with RRR ranging from 3.0 to 39.3. In the RRR = 3.0 sample, a single peak is observed at $T_{CDW}$. In contrast, the RRR = 10.8 exhibits two distinct peaks, and with further increasing RRR each of these features splits into additional peaks. Remarkably, the RRR = 30.0 sample shows two to three peaks within each of the two clusters, indicating that more than two phase transitions occur in the immediate vicinity of $T_{CDW}$ in the cleaner crystals. To confirm that this splitting reflects an intrinsic property of the sample rather than spatial inhomogeneity, we measured the elastocaloric effect ($\frac{dT}{d\epsilon}$), a bulk thermodynamic probe that is proportional to the heat capacity anomaly across a phase transition~\cite{ECE_Ikeda}. The bottom panel of Fig.~\ref{fig:RRR_phasetransitions}b shows $\frac{dT}{d\epsilon}$ for the RRR = 30.0 sample  used for the resistivity in the top panel. Two clusters of peaks are resolved in $\frac{dT}{d\epsilon}$, each containing up to three features, in excellent agreement with $\frac{d\rho_{xx}}{dT}$ and providing thermodynamic evidence that the splitting of the CDW transitions in high-RRR samples is an intrinsic phenomenon.

The RRR dependence of transition temperature and their splitting, collected from nearly 40 samples of \svs, is summarized in Fig.~\ref{fig:RRR_phasetransitions}c. Green points denote samples showing a single peak in $\frac{d\rho_{xx}}{dT}$, orange points denote samples showing multiple peaks in $\frac{d\rho_{xx}}{dT}$ and the vertical bars indicate the spread of the multiple transition temperatures. The low-RRR regime is shown in greater detail in panel d, where a sharp boundary near RRR = 4 separates the single-transition and multi-transition regimes. A previous report identified the splitting of the CDW transition in a single sample of \svs\ with RRR = 9, accompanied by the pronounced in-plane resistivity anisotropy between the two transitions~\cite{intermediatenematic_SVS}. Our measurements extend this observation to a systematic study across nearly 40 samples, establishing that the splitting is a generic feature of the disorder-tuned phase diagram rather than an isolated observation, and revealing additional substructure that emerges only in the highest-quality samples.

% Similarly, Fig.~\ref{fig:RRR_phasetransitions}b shows $\rho_{zz}$ normalized by the \SI{300}{K} value as a function of temperature for samples of \svs\ with $RRR_{zz}$ being tuned by several times. A drop in $\rho_{zz}$ is again observed upon cooling through $T_{CDW}$. Like in the $\rho_{xx}$ samples, a significant change in the high temperature $\rho_{zz}$ is observed with increasing $RRR_{zz}$. This explains previous discrepancies in literature reports of $\rho_{zz}$ at high temperatures~\cite{SVS_PG, universalsublinearresistivity_SVS_CVS}.

\begin{figure*}
    \centering
    \includegraphics[width=0.9\textwidth]{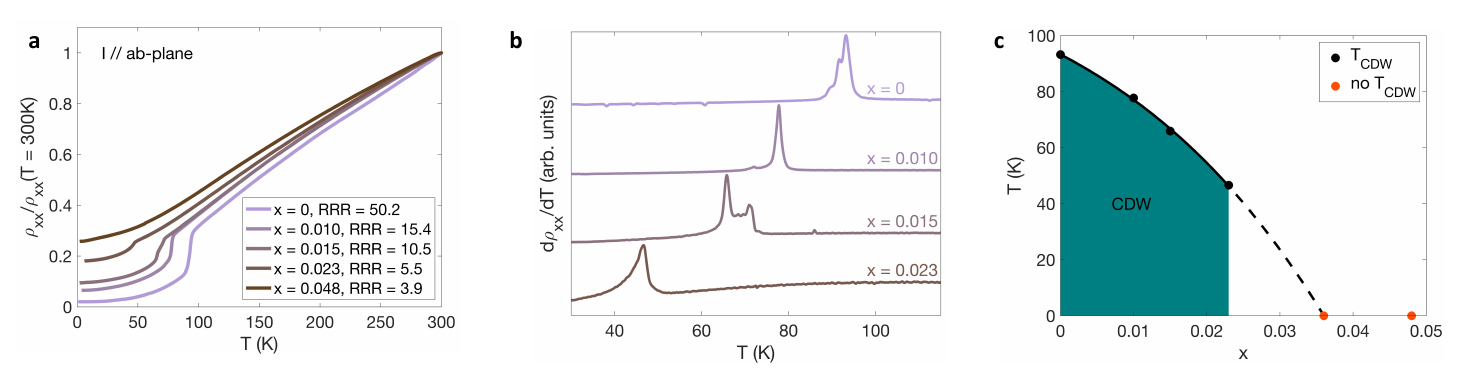}
    \caption{\textbf{Phase diagram of \svcs.} \textbf{a,} In-plane resistivity $\rho_{xx}$, normalized by its \SI{300}{K} value, as a function of temperature for \svcs\ samples with $x$ ranging from 0 to 0.048. RRR decreases with increasing $x$ as substitutional disorder is introduced. \textbf{b,} $d\rho_{xx}/dT$ for several \svcs\ samples, offset vertically for clarity. The low-$x$ (high-RRR) samples retain the multi-peak structure characteristic of clean \svs, whereas the $x = 0.023$ sample exhibits only a single feature, mirroring the low RRR sample in the H$_2$-cleaning series. $T_{CDW}$ is progressively suppressed with increasing $x$. \textbf{c,} Temperature--composition phase diagram constructed from the data in panels a and b; only the onset $T_{CDW}$ is shown. Black points mark samples with a CDW transition; red points mark samples in which no transition is observed. The dashed line is a guide to the eye extrapolating to the critical composition at which the CDW is fully suppressed.}
    \label{fig:SVCS_phasediagram}
\end{figure*}

To complement the disorder tuning approach based on H$_2$-cleaning of the starting vanadium, we grew single crystals of \svcs\ by introducing Cr into the same recipe used to produce the highest RRR \svs\ samples. Cr substitution serves a dual purpose: it dopes the system (nominally with electrons), as confirmed by the evolution of the quantum oscillation frequency in Sec.~\ref{sec:QOs}, while simultaneously introducing substitutional disorder that reduces the carrier mobility. Fig.~\ref{fig:SVCS_phasediagram}a presents $\rho_{xx}$ normalized by its \SI{300}{K} value as a function of temperature for samples of \svcs\ with various $x$. As expected, RRR decreases monotonically with increasing $x$. The corresponding $\frac{d\rho_{xx}}{dT}$ curves, shown in Fig.~\ref{fig:SVCS_phasediagram}b, reveal that $T_{CDW}$ is progressively suppressed with increasing $x$, in agreement with previous reports, ~\cite{natureofCDW_SVS, tuningSVS_Crdoping}. Importantly, the low $x$ (and correspondingly high RRR) samples retain the multi-transition structure observed in pure \svs , whereas the $x = 0.023$ (low RRR) sample exhibits only a single feature in $\frac{d\rho_{xx}}{dT}$. This parallel behavior across two independent disorder axes---H$_2$ cleaning of the V starting material and Cr substitution on the V site---further supports that the multi-transition structure is intrinsic to clean \svs\ and is smeared by disorder, regardless of the microscopic origin of that disorder. The $x-T$ phase diagram constructed from these data is summarized in Fig.~\ref{fig:SVCS_phasediagram}c; for simplicity only the onset of $T_{CDW}$ is shown. The suppression of $T_{CDW}$ with Cr doping has been attributed to the reduction of the c-axis lattice constant with increasing $x$~\cite{natureofCDW_SVS} in agreement with the ``rattling chain" model~\cite{rattlingchain}. 

% Similarly, Fig.~\ref{fig:Crphasediagram}b shows $\rho_{zz}$ normalized by the \SI{300}{K} value as a function of temperature for samples of \svcs\ with different $x$ where it can be seen that $T_{CDW}$ is suppressed with increasing $x$.
\subsection{Symmetry-resolved elastoresistivity}

\begin{figure*}
    \centering
    \includegraphics[width=0.9\textwidth]{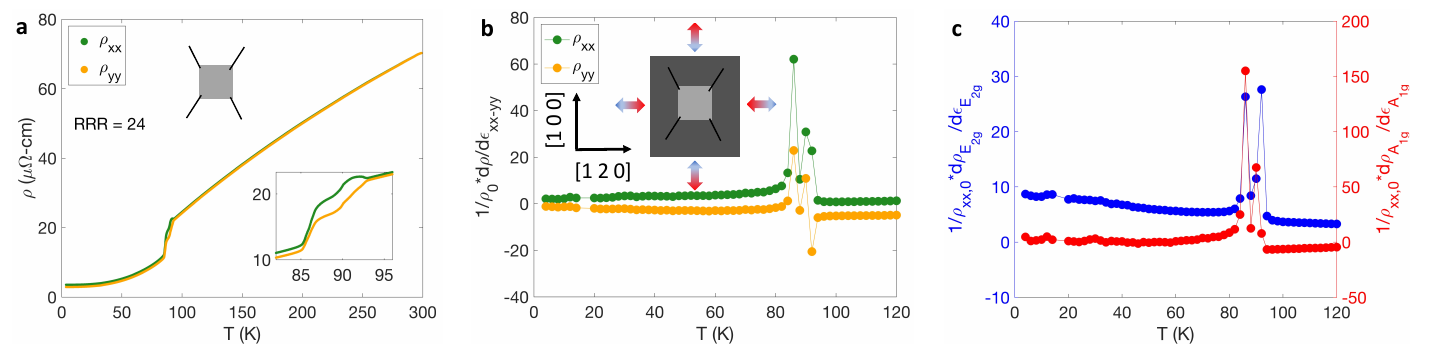}
    \caption{\textbf{Elastoresistivity of the intermediate phase in \svs} \textbf{a,} In-plane resistivity $\rho_{xx}$ and $\rho_{yy}$ as a function of temperature, measured simultaneously in a Montgomery geometry of a free-standing \svs\ crystal with a RRR of 24. Inset: Zoom-in near the transitions, where $\rho_{xx}$ and $\rho_{yy}$ split between the upper and lower transitions and recover below them. \textbf{b,} Normalized strain derivatives $(1/\rho_0)d\rho_{xx}/d\epsilon_{xx-yy}$ and $(1/\rho_0)d\rho_{yy}/d\epsilon_{xx-yy}$ as a function of temperature, exhibiting sharp features within the intermediate phase. Inset: sample orientation, with anisotropic strain applied along $[\overline{1}20]$ and $[100]$ directions \textbf{c,} The $E_{2g}$ (blue, left axis) and $A_{1g}$ (red, right axis) components of the elastoresistivity, obtained from the antisymmetric and symmetric combinations of the channels in (b), respectively. Both components are sharply enhanced within the intermediate phase and featureless outside it.}
    \label{fig:elastoresistivity}
\end{figure*}

The multi-transition structure resolved in Sec.~\ref{sub:transitions} raises the question of which symmetries are broken at each transition. Of particular interest is the intermediate phase between the upper and lower transitions, which a prior study reported to exhibit pronounced in-plane resistivity anisotropy~\cite{intermediatenematic_SVS}, suggesting a possible breaking of rotational symmetry. Here we use symmetry-resolved elastoresistivity---a bulk thermodynamic probe sensitive to the rotational symmetry of the electronic state---to directly identify the broken symmetry of the intermediate phase. We focus on a sample with RRR~=~24, cut into a square with edges along $[120]$ and $[100]$ with contacts in a Montgomery geometry enabling simultaneous measurement of $\rho_{xx}$ and $\rho_{yy}$. Fig.~\ref{fig:elastoresistivity}a presents $\rho_{xx}$ and $\rho_{yy}$ as a function of temperature; a clear splitting between the two channels emerges between the upper and lower transitions and recovers below them, consistent with the anisotropy reported in Ref.~\cite{intermediatenematic_SVS}. The sample was then mounted on a piezoelectric strain cell allowing strain to be applied in the xy-plane. Fig.~\ref{fig:elastoresistivity}b shows the normalized (by $\rho_0$, the zero-strain resistivity) strain derivatives $d\rho_{xx}/d\epsilon_{xx-yy}$ and $d\rho_{yy}/d\epsilon_{xx-yy}$, which exhibit sharp features of opposite sign within the same temperature window. The antisymmetric and symmetric combinations of these channels isolate the $E_{2g}$ and $A_{1g}$ components of the elastoresistivity~\cite{shapiroMeasurementB1gB2g2016}, presented in Fig.~\ref{fig:elastoresistivity}c. Sharp features in both channels align one-to-one with the peaks in $d\rho_{xx}/dT$, and both responses are confined to the intermediate window and vanish outside it.

% mounted on a piezoelectric strain cell with contacts in the Montgomery geometry, enabling simultaneous measurement of $\rho_{xx}$ and $\rho_{yy}$ under uniaxial strain applied along $[\overline{1}20]$.

The pronounced enhancement of the $E_{2g}$ elastoresistivity within this window indicates that the resistivity is extremely susceptible to anisotropic strain, as expected for the strain-induced detwinning of nematic domains in the intermediate phase. The even larger $A_{1g}$ response arises from two sources: an admixture of $E_{2g}$ and $A_{1g}$ irreducible representations once three-fold symmetry is spontaneously broken, and a heat-capacity-like contribution $\propto (d\rho/dT)(dT_c/d\epsilon_{A_{1g}})$ that is generic to any sharp resistive anomaly~\cite{ECE_Ikeda}. The absence of any $E_{2g}$ or $A_{1g}$ signal outside the intermediate window places a stringent upper bound on additional rotational-symmetry-breaking transitions in the bulk. This is in apparent tension with a recent scanning tunneling microscopy study reporting intra-unit-cell nematic order and signatures of a Pomeranchuk instability extending well beyond the immediate vicinity of $T_{CDW}$~\cite{hasan_svsnematic}. While we cannot exclude surface-confined nematicity, our bulk transport data show no evidence for an electronically driven nematic instability outside the narrow window between the upper and lower CDW transitions. 

\subsection{Magnetoresistance and quantum oscillations}
\label{sec:QOs}

\begin{figure*}
    \centering
    \includegraphics[width=0.9\textwidth]{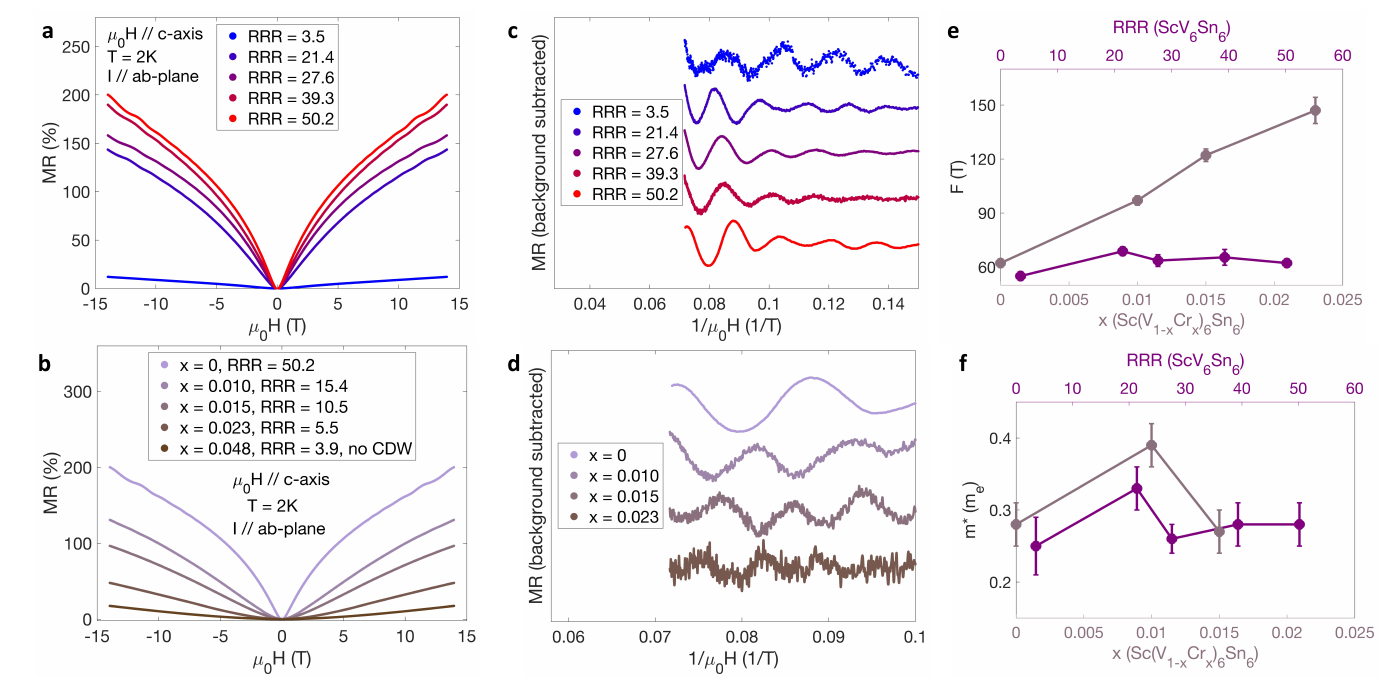}
    \caption{\textbf{Magnetoresistance and quantum oscillations of \svs\ and \svcs.} \textbf{a,} In-plane magnetoresistance (MR) at \SI{2}{K} as a function of $\mu_0H$ applied along the c-axis, for samples of \svs\ with varying RRR. The MR increases by roughly an order of magnitude from the lowest- to highest-RRR sample, and quantum oscillations are visible by eye in the cleanest samples.\textbf{b,} Corresponding MR of \svcs\ for varying $x$; the MR decreases rapidly with increasing $x$ as substitutional disorder reduces the mobility. \textbf{c,} Background subtracted MR of the \svs\ samples versus inverse field $\frac{1}{\mu_0H}$, offset vertically for clarity. \textbf{d,} Same as (c) for the \svcs\ samples. \textbf{e,} Quantum oscillation frequency $F$ for \svs\ versus RRR  (top x-axis, purple) and for \svcs\ versus $x$  (bottom x-axis, gray). \textbf{f,} Effective mass extracted from the temperature dependence of the oscillation amplitude, plotted using the same axes and colors as (e). $m^*$ shows no substantial change with either RRR or $x$, indicating that the band structure of this pocket is unaffected by disorder or Cr substitution.}
    \label{fig:MRxx}
\end{figure*}

The electronic structure of \svs\ has been predicted by first-principles calculations to host several small pockets near the Brillouin zone boundary~\cite{quantumoscillationsevidence_SVS,SVS_PG}. Here we directly probe these pockets through quantum oscillations to characterize their evolution with disorder and Cr substitution. The oscillations are observed in the in-plane magnetoresistance, MR\,(\%) = $[\rho(\mu_0 H) - \rho(0)]/\rho(0) \times 100$\%, with the magnetic field applied along the $c$ axis. Fig.~\ref{fig:MRxx}a presents the MR of \svs\ at \SI{2}{K} for samples with varying in-plane RRR. As expected for systems with increased carrier mobility, higher-RRR samples display a markedly larger MR; the highest-RRR sample reaches an MR roughly an order of magnitude larger than that of the lowest-RRR sample. Quantum oscillations are clearly resolved by eye in the higher-RRR traces. At the highest applied fields, the MR of the cleanest samples becomes visibly sub-linear, indicating the onset of partial saturation associated with the high carrier mobility. The corresponding MR of \svcs, presented in Fig.~\ref{fig:MRxx}b, decreases rapidly with increasing $x$, again consistent with the reduction of $\mu$ by substitutional disorder.

Subtracting a polynomial background from the MR presented in panels a and b and plotting these data against inverse magnetic field reveals a single dominant quantum oscillation in both \svs\ and \svcs\ as shown in panels c and d. This is supported by fast Fourier transforms (FFTs) of the data presented in Fig.~\ref{fig:MRxx}c which show one quantum oscillation frequency in \svs, in good agreement with previous reports~\cite{quantumoscillationsevidence_SVS}. This procedure also shows one quantum oscillation frequency in low-doped \svcs. To our knowledge, quantum oscillations in \svcs\ have not been reported previously; their observation here is enabled by the high sample quality achieved through our growth procedure. The  extracted frequency at \SI{2}{K}, obtained from standard Landau level indexing, are shown in Fig.~\ref{fig:MRxx}e in purple for \svs\ as a function of RRR (top x-axis) and in gray for \svcs\ as a function of $x$ (bottom x-axis). For \svs\, the frequency remians essentially constant at $\approx$\SI{60}{T} across the full RRR range, indicating that the Fermi energy ($E_F$) is unchanged by disorder. In contrast, for \svcs\ the frequency more than doubles upon only 2\% Cr substitution on the V site. While Cr-doping nominally electron dopes this system, the prediction of both small hole and electron pockets does not allow us to uniquely identify this pocket as electron-like . 

The corresponding effective mass ($m^*$), expressed in units of the free electron mass ($m_e$), are presented in Fig.~\ref{fig:MRxx}f using the same conventions as panel e. The effective masses were extracted by fitting the temperature dependence of the FFT peak amplitude to the damping factor ($R_T$) of the Lifshitz-Kosevich formula,

\begin{equation}
    R_T\propto\frac{\alpha (m^*/m_e)T/\mu_0H_{avg}}{\sinh(\alpha (m^*/m_e)T/\mu_0H_{avg})}
    \label{eqn:tempdepQOs}
\end{equation}

where $\mu_0H_{avg}$ is the average of the magnetic field used in calculating the FFT and $\alpha \approx 14.69$T/K. No substantial change in $m^*$ is observed increasing RRR for \svs\ nor with increasing $x$ in \svcs\, indicating that neither disorder nor Cr substitution significant alters the underlying band structure of the small pocket. Having established the existence and stability of this small, high-mobility pocket across the full disorder range, we now turn to its consequences for the magnetotransport, where it plays a central role in the Hall response.

\subsection{Hall effect}

\begin{figure*}
    \centering
    \includegraphics[width=0.9\textwidth]{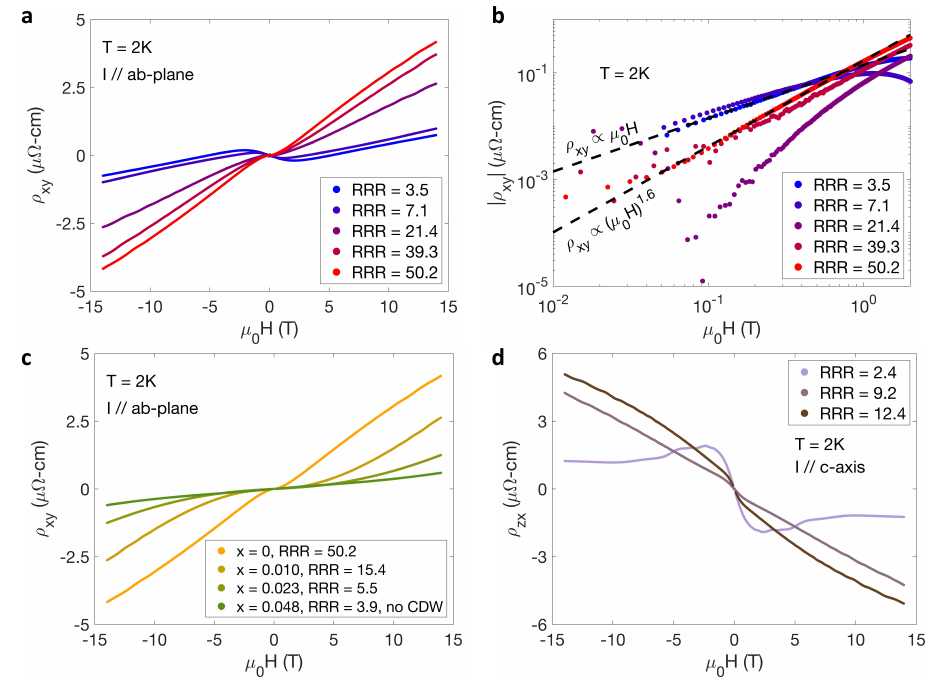}
    \caption{\textbf{Multi-band Hall effect of \svs\ and \svcs.} \textbf{a,} Hall resistivity $\rho_{xy}$ at \SI{2}{K} as a function of $\mu_0H$ for \svs\ samples with a variety of in-plane RRR. \textbf{b,} The same data on a log-log scale. In the lowest-RRR sample, $\rho_{xy}$ exhibits a clear linear-in-field regime at low fields (dashed line, $\rho_{xy} \propto \mu_0 H$), the hallmark of the weak-field Hall effect. This weak-field regime progressively shrinks with increasing RRR and is no longer observable within our field range in the highest-RRR sample, consistent with the dramatic increase in mobility. \textbf{c,} $\rho_{xy}$ at \SI{2}{K} for \svcs\ samples with varying $x$. The low-field non-linearity shifts to higher field scales with increasing $x$, consistent with the reduction of mobility by substitutional disorder. \textbf{d,} Out-of-plane $\rho_{zx}$ ($I \parallel c$) at \SI{2}{K} for \svs\ samples with varying out-of-plane RRR. The plateau-like feature in the lowest-RRR sample, on which a previous AHE interpretation was based, is absent in cleaner samples---opposite to the expectation for an intrinsic anomalous Hall effect.}
    \label{fig:Hall}
\end{figure*}

% Now that we have firmly established \svs\ as a system with multiple small, light electron pockets, we will examine the low temperature Hall effect in both \svs\ and \svcs\ which are best explained by a multi-band system without an AHE.

Having characterized the small, high-mobility pocket through quantum oscillations, we now turn to its consequences for the Hall response. Fig.~\ref{fig:Hall}a presents Hall resistivity ($\rho_{xy}$) at \SI{2}{K} as a function of magnetic field for \svs\ samples spanning a wide range of RRR. A pronounced nonlinear $\rho_{xy}$ is observed in the low-field regime, in agreement with previous reports~\cite{universalsublinearresistivity_SVS_CVS, SVS_PG}. Such a non-linear Hall effect is generally interpreted in one of two ways: as a signature of multi-band transport, particularly when high-mobility small Fermi pockets coexist with low-mobility large Fermi pockets, or as an anomalous Hall effect (AHE) arising from time-reversal symmetry breaking. In a multi-band system, the field scale at which the Hall response crosses over from low-field to high-field behavior is set by the inverse of the highest mobility. The field scale of non-linearity in $\rho_{xy}$ decreases with increasing RRR (and correspondingly increasing mobility $\mu$), a behavior characteristic of an ordinary multi-band Hall effect. This trend is shown most clearly on a logarithmic field scale in Fig.~\ref{fig:Hall}b. In the lowest-RRR sample, $\rho_{xy}$ displays a well-defined linear-in-field regime at low fields, $\rho_{xy} \propto \mu_0 H$, corresponding to the weak-field Hall effect in which $\mu_0 H \ll \mu^{-1}$ for all carriers. As RRR increases, this weak-field regime systematically shrinks: the crossover to non-linear behavior moves to progressively lower fields, until in the highest-RRR samples no linear regime is observable. This is precisely the behavior expected for an ordinary multi-band Hall effect, in which the weak-field boundary is set by $\mu_\text{max}^{-1}$ and therefore shrinks as the carrier mobility increases.

To make this argument quantitative, we fit all curves to a standard two-band model following the methodology of Ref.~\cite{twoband_coey}; the complete set of fit parameters is presented in Appendix A. While the fitted electron and hole mobilities increase systematically with RRR, which is expected for samples with reduced disorder, the extracted electron carrier density simultaneously increases by several orders of magnitude with across the series, driven by the increase in the high-field slope of $\rho_{xy}$ with increasing RRR. As established in Sec.~\ref{sec:QOs}, the $E_F$ does not change appreciably across the same RRR range, and the carrier density should not either. This internal inconsistency unambiguously demonstrates that the standard two-band picture fails to describe the Hall response of \svs. Although a similar limitation of the two-band model has been noted previously in the CDW phase~\cite{SVS_PG}, the fit quality gets progressively worse with increasing RRR (Appendix A), pointing to a multi-band electronic landscape with several small, high-mobility pockets that the two-band approximation cannot accommodate.

% Fig.~\ref{fig:Hall}b displays $\Delta\rho_{xy} = \rho_{xy}-\rho_{xy}^{fit}$ for the two-band model at \SI{2}{K} for low and high RRR samples of \svs\ and it can be seen that the residuals increase in size with increasing RRR in the low-field regime.

% This is potentially indicative of the small electron pockets playing more of a role in electronic transport in higher mobility samples.

A complementary piece of evidence against the two-band interpretation comes from Cr doping series. Fig.~\ref{fig:Hall}c presents $\rho_{xy}$ at \SI{2}{K} as a function of $\mu_0H$ for samples of \svcs\ with various $x$. With increasing $x$ (and correspondingly decreasing $\mu$), the low-field non-linearity systematically shifts to higher field scales, consistent with an ordinary multi-band Hall effect. Furthermore, the high-field slope of $\rho_{xy}$ decreases with increasing $x$, which within a two-band picture would imply that Cr substitution hole dopes the system. This conclusion contradicts the expected electron doping from Cr substitution. The most natural explanation is that \SI{14}{T} lies well below the high-field limit of the system, so that the apparent ``high-field'' slope of $\rho_{xy}$ has not yet reached its asymptotic value.  Robust extraction of a field-induced AHE component therefore requires reaching a true high-field saturation, which our data demonstrate is not accessible within \SI{14}{T}.

It has also been conjectured that $\rho_{zx}$ in \svs\ displays an AHE component~\cite{QuantumOscillationsRevealing_SVS, universalsublinearresistivity_SVS_CVS}, an argument based on the field dependence of $\rho_{zx}$ at low temperatures in samples with relatively low out-of-plane RRR (RRR$_{zz}$, defined as $\rho_{zz}(300\,\text{K})/\rho_{zz}(3.5\,\text{K})$). To test this interpretation, we measured $\rho_{zx}$ across a range of out-of-plane RRR; the results at \SI{2}{K} are shown in Fig.~\ref{fig:Hall}d. The field dependence of $\rho_{zx}$ evolves dramatically with increasing out-of-plane RRR, and the field scale of low-field anomaly is rapidly reduced in cleaner samples,  which is the same mobility-dependent crossover signature identified in $\rho_{xy}$. Notably, the plateau-like feature in the low-RRR sample on which the AHE interpretation was based is entirely absent in higher-RRR samples. This is exactly the opposite of the expectation for an intrinsic AHE, which should become sharper, not weaker, as disorder is reduced. This RRR-dependent evolution is instead the hallmark of an ordinary multi-band Hall response.

\section{Discussion}

The systematic mobility tuning achieved in this work, with RRR spanning more than an order of magnitude and exceeding the previous record by a factor of five, has enabled three principal findings, which we summarize and place in context here.

First, our quantum oscillation measurements identify a small, light pocket whose frequency ($\approx$\SI{60}{T}) and effective mass ($m^* \approx 0.3\,m_e$) are unchanged across the full RRR range of \svs, and whose frequency more than doubles upon $\approx$2\% Cr substitution. These observations are in good agreement with first-principles calculations~\cite{SVS_PG} and establish that disorder in our growth procedure modifies the carrier mobility while leaving the underlying band structure intact. This validates the comparison of samples with different RRR as a clean mobility-tuning axis, and underpins the analysis in the remainder of the paper.

Second, the field scale, magnitude, and shape of the Hall anomalies in both $\rho_{xy}$ and $\rho_{zx}$ are systematically controlled by the carrier mobility, established through three mutually consistent observations: the field scale of the low-field non-linearity in $\rho_{xy}$ shifts to lower values with increasing RRR; a two-band fit to $\rho_{xy}$ requires an unphysical RRR-dependence of the carrier density, in direct conflict with the fixed Fermi energy established by quantum oscillations; and the plateau-like feature in $\rho_{zx}$ that originally motivated the AHE interpretation~\cite{QuantumOscillationsRevealing_SVS} is absent in cleaner samples, the opposite of the expectation for an intrinsic AHE. Taken together, these results indicate that the Hall anomalies in \svs\ originate from the small, high-mobility pockets revealed by our quantum oscillation data rather than from a time-reversal-symmetry-breaking state. This conclusion mirrors the recently established situation in \ch{CsV3Sb5}~\cite{CVS_tinypockets} and suggests a common phenomenology across the kagome CDW family. More broadly, these results highlight the importance of mobility tuning as a diagnostic for separating intrinsic anomalous Hall contributions from ordinary multi-band effects in metallic systems with significantly disparate Fermi pockets. Moreover, these results are in good agreement with a finding of a lack of TRS breaking from high-resolution polar Kerr measurements~\cite{kapitulnik_SVS}.

Third, the CDW transition in clean \svs\ is not a single transition but a sequence of closely spaced transitions that emerges sharply above a threshold of RRR $\approx 4$. In samples with RRR $\gtrsim 20$, as many as six distinct features are resolved within a few Kelvin of $T_{CDW}$, and the elastoresistivity demonstrates that the intermediate phase between the upper and lower transitions breaks the three-fold rotational symmetry of the parent lattice. The agreement between the resistivity, elastocaloric, and elastoresistivity data confirms that this multi-transition structure is intrinsic and thermodynamic in origin, while the parallel behavior observed across two independent disorder axes (H$_2$ cleaning and Cr substitution) demonstrates that the splitting is a generic property of clean \svs\ that is smeared by disorder regardless of its microscopic origin. The microscopic nature of the additional substructure resolved in our highest-RRR samples beyond the single intermediate phase identified in Ref.~\cite{intermediatenematic_SVS} remains an open question. 

In summary, by establishing a growth protocol that produces \svs\ crystals with mobility tunable over more than an order of magnitude, we have shown that (i) the Hall anomalies previously attributed to an anomalous Hall effect originate from ordinary multi-band transport, (ii) the CDW transition in clean \svs\ consists of a sequence of closely spaced symmetry-breaking transitions, and (iii) the intermediate phase is a bulk nematic state that breaks the three-fold rotational symmetry of the hexagonal lattice. These results refine the symmetry-breaking landscape of \svs\ and motivate further symmetry-resolved investigations of the kagome CDW family.

\section{Methods}
\label{sec:methods}

Single crystals of \svs\ and \svcs\ were grown utilizing a flux method similar to those previously reported for \svs~\cite{svs_discovery}. 15 gram mixtures of Sc pieces (99.9\%), V pieces (99.9\%), Cr powder (99.95\%), and Sn shot (99.999\%) were loaded and subsequently vacuum sealed in quartz tubes with atomic ratios Sc:(V,Cr):Sn 1:3:80. These were heated up to 1150$^\circ$C in 12 hours, held at this temperature for 12 hours, then cooled to 700$^\circ$C in 120 hours. Then the growths were decanted in a centrifuge to separate the excess flux from the crystals. Some of the V pieces were cleaned by being heated to 800$^\circ$C for about 12 hours in flowing Ar/\ch{H2} gas (5\% molar concentration of hydrogen) and were then stored in an Ar glovebox. The highest RRR \svs\ and all of the \svcs\ samples were synthesized with this clean V. For the other \svs\ growths, cleaned V pieces were mixed with as-received V pieces in varying ratios. This allowed for a relatively smooth tuning of crystal quality. Some of the crystals grown in this way had several secondary phases with superconducting transitions embedded in them including Sn and \ch{V3Si}. Samples were carefully polished and cut to attempt to exclude these phases, and samples which showed significant superconducting transitions in transport measurements are not reported.

The doping concentration \textit{x} of each sample of \svcs\ used for measurements was determined using energy-dispersive X-ray spectroscopy (EDX) with a Sirion XL30 scanning electron microscope. Each crystal was polished prior to EDX measurements to remove residual flux on the surface of the crystal, and at least 8 points on each crystal were measured. The measured spread in \textit{x}$_{EDX}$ for each crystal used in this study was limited, indicating the doping is reasonably homogeneous throughout the crystal. Throughout this paper the measured \textit{x}$_{EDX}$ is referred to as \textit{x} for simplicity. Some samples of \svs\ had EDX measurements performed on them and the results are shown in Appendix B; within the error bars of EDX no conclusions can be made relating the stoichiometry of \svs\ to $RRR_{xx}$.

Transport measurements were performed on samples that were polished and cut by a wire saw to be bars with dimensions roughly \SI{1}{mm} $\times$ \SI{0.4}{mm} $\times$ \SI{0.05}{mm}. Silver paste and gold wires were used to make 5 point (Hall pattern) contacts. These measurements were performed in a Quantum Design Dynacool Physical Property Measurement System (PPMS) with standard lock-in techniques in temperatures ranging from \SI{2}{K} to \SI{300}{K} and in magnetic fields up to \SI{14}{T}. To eliminate any contributions from contact misalignment the in-line and Hall resistivities were symmetrized and anti-symmetrized with magnetic field, respectively.

Elastocaloric measurements were performed by gluing a sample previously used for transport measurements across the gap of a home-made three piezoelectric strain cell, modeled after the commercial Razorbill CS-100 strain cell. The sample was secured to the titanium cell across a gap of \SI{0.5}{mm} using Stycast 2850FT Epoxy. An AC voltage of \SI{2.5}{V} root mean square at \SI{23}{Hz} was applied to the outer piezoelectric stacks of the strain cell, which corresponded to creating an AC displacement of the sample of approximately 0.005\% of its length. This frequency was experimentally determined by measuring the elastocaloric signal at \SI{100}{K} for frequencies in the range of 10-\SI{100}{Hz}, and choosing the frequency with the largest response. This frequency was at the peak of the relevant thermal transfer function, which did not observably shift in the temperature range measured~\cite{ECE_Ikeda}. Due to the brittle nature of \svs, we found that even small tensile strains broke the samples. As such, a small compressive strain is applied to the sample during cooling. This increased each of the phase transition temperatures, in agreement with Ref.~\cite{svs_strain_Tcdwincrease} which found that compressive in-plane strain increased $T_{CDW}$. %DC voltages were applied to the inner PZT to reach a strain range of 0.7\%. To approximate the strain the sample experienced a capacitor built in the strain cell was measured to provide the relative displacement of the sample plates, which was divided by the length of the gap. This however only approximated $\varepsilon_{xx}$ of the sample as it assumes a 100\% strain transmission. Strain transmission at high temperatures (350K) decreases due to the softening of the epoxy used, but this was taken into account by measuring a strain gauge with a constant gauge factor glued an identical way as a calibration.

Elastoresistivity measurements were performed on samples that were polished and cut to be squares approximately \SI{0.6}{mm} $\times$ \SI{0.6}{mm} $\times$ \SI{0.05}{mm}. The samples were glued onto PICeramic 5x5x9 piezostacks using the previously mentioned epoxy. Contacts were arranged at the corners of the square to perform modified Montgomery method measurements, in which both $\rho_{xx}$ and $\rho_{yy}$ were obtained via resistance measurements in directions parallel and perpendicular to the poling direction of the piezoelectric stack (see Ref.~\cite{modifiedmontgomery}). The sample and stack would be stabilized at a set temperature, and a voltage range would be applied to the piezo stack such that the slopes of both the symmetric and antisymmetric components of the resistivity tensor ($\rho_{E_{2g}}$ and $\rho_{A_{1g}}$ respectively) against strain could be determined. 
% The in-plane orientation of several crystals of \svs\ and \svcs\ were determined using a Rigaku Mini-Flex 600 system with a Cu source. It was determined that the crystal facets aligned with the $[1\,0\,0]$ ($[2\,\overline{1}\,\overline{1}\,0]$) direction in agreement with the findings of Ref.~\cite{QuantumOscillationsRevealing_SVS}.

\section*{Acknowledgments}

 This work was primarily supported by NSF MRSEC at UW (DMR-2308979). This material is based upon work supported by the National Science Foundation Graduate Research Fellowship Program under Grant No. DGE-2140004. Any opinions, findings, and conclusions or recommendations expressed in this material are those of the authors and do not necessarily reflect the views of the National Science Foundation. Part of this work was conducted at the Molecular Analysis Facility, a National Nanotechnology Coordinated Infrastructure site at the University of Washington which is supported in part by the National Science Foundation (grant NNCI-1542101), the University of Washington, the Molecular Engineering \& Sciences Institute, the Clean Energy Institute, and the National Institutes of Health.

 \section*{Author Contributions}
 J.M.D., E.R., and C.H. grew the samples. J.M.D. and E.R. performed the measurements. J.-H.C. and X.X. oversaw the project. J.M.D. and J.-H.C. wrote the manuscript with input from all authors.

 \section*{Appendix A: Two-Band Hall Fitting}

 \begin{figure*}
    \centering
    \includegraphics[width=0.9\textwidth]{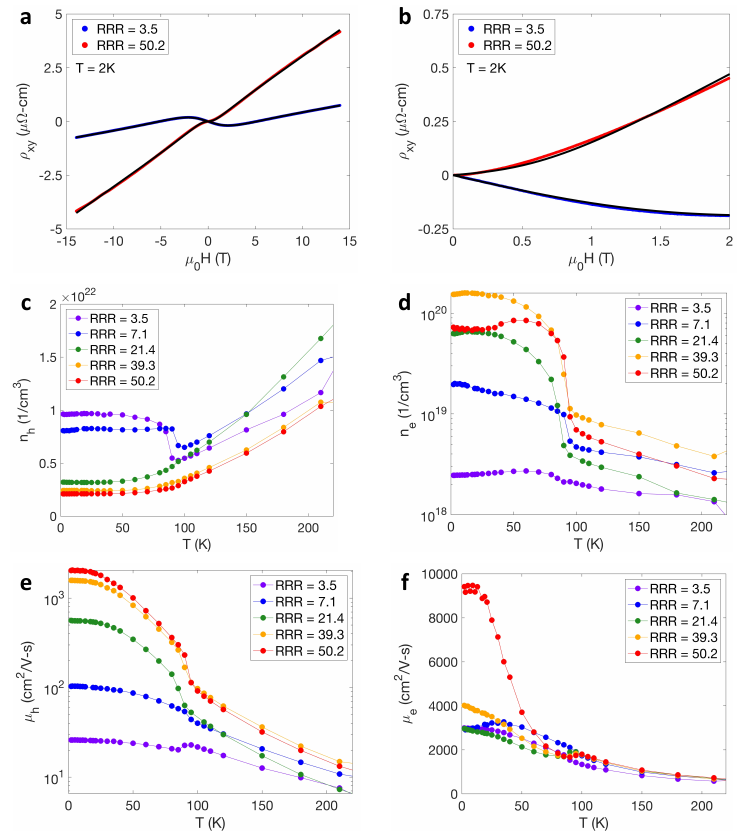}
    \caption{\textbf{Two-band Hall fitting.} \textbf{a,} $\rho_{xy}(\mu_0H)$ at \SI{2}{K} for samples of \svs\ with RRRs of 3.5 (blue) and 50.2 (red). Black lines indicate fits to the two-band formula from Ref.~\cite{twoband_coey}. \textbf{b,} The same data as in panel a focused on the low-field regime. Note that the low RRR sample data can be fit better than the high RRR sample data. \textbf{c, d, e, f,} $n_h$, $n_e$, $\mu_h$, and $\mu_e$ as a function of temperature extracted from two-band Hall fitting for samples with varied RRR.}
    \label{fig:twobandHallfits}
\end{figure*}

Fig.~\ref{fig:twobandHallfits}a presents $\rho_{xy}(\mu_0H)$ at \SI{2}{K} for samples of \svs\ with RRRs of 3.5 (blue) and 50.2 (red). Black lines indicate fits to the two-band formula from Ref.~\cite{twoband_coey}. While these fits generally capture the shape of these curves, it can be seen that the fit quality degrades (particularly in the low-field regime) for high RRR samples as shown in panel b. Panels c, d, e, and f show $n_h$, $n_e$, $\mu_h$, and $\mu_e$ respectively as a function of temperature extracted from two-band Hall fitting for samples with varied RRR. While the increase in mobilities observed with increasing RRR is to be expected from increasing the crystal quality of the samples, the apparent increase (decrease) in the electron (hole) concentrations is surprising as described in the main text.

\section*{Appendix B: EDX}

 \begin{figure*}
    \centering
    \includegraphics[width=0.9\textwidth]{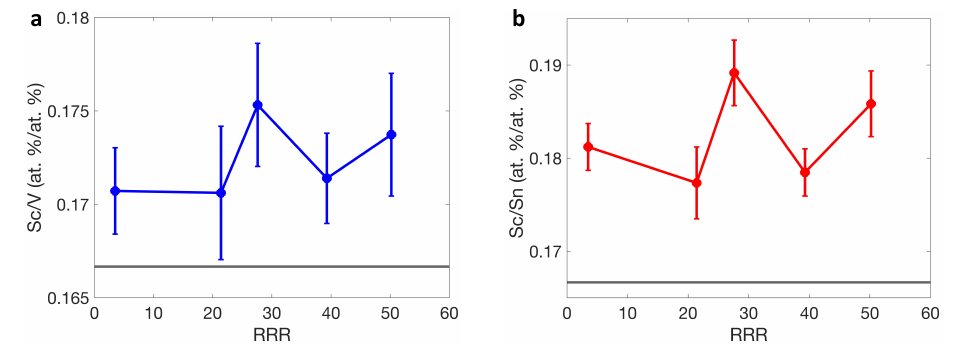}
    \caption{\textbf{EDX results.} \textbf{a,} Sc/V ratio plotted as a function of RRR. \textbf{b,} Sc/Sn ratio plotted as a function of RRR.}
    \label{fig:edx}
\end{figure*}

Fig.~\ref{fig:edx} presents EDX results on samples of \svs\ with varying RRRs. Black lines indicate the ``ideal" 1:6 atomic ratio in both plots. Neither the Sc/V ratio nor the Sc/Sn ratio changes systematically as a function of RRR.

\clearpage
\newpage

\bibliography{main}

@article{Ortiz2020,
  title = {\ch{CsV3Sb5}: A ${{Z}_{2}}$ Topological Kagome Metal with a Superconducting Ground State},
  author = {Ortiz, Brenden R. and Teicher, Samuel M. L. and Hu, Yong and Zuo, Julia L. and Sarte, Paul M. and Schueller, Emily C. and Abeykoon, A. M. Milinda and Krogstad, Matthew J. and Rosenkranz, Stephan and Osborn, Raymond and Seshadri, Ram and Balents, Leon and He, Junfeng and Wilson, Stephen D.},
  journal = {Phys. Rev. Lett.},
  volume = {125},
  issue = {24},
  pages = {247002},
  numpages = {6},
  year = {2020},
  month = {Dec},
  publisher = {American Physical Society},
  doi = {10.1103/PhysRevLett.125.247002},
  url = {https://link.aps.org/doi/10.1103/PhysRevLett.125.247002}
}

@article{Jiangetal2021,
  author    = {Jiang, Yu-Xiao and Yin, Jia-Xin and Denner, M. Michael and Shumiya, Nana and Ortiz, Brenden R. and Xu, Gang and Guguchia, Zurab and He, Junyi and Hossain, Md Shafayat and Liu, Xiaoxiong and Ruff, Jacob and Kautzsch, Linus ... and Hasan, M. Zahid and Wilson, Stephen D.},
  title     = {Unconventional chiral charge order in kagome superconductor \ch{KV3Sb5}},
  journal   = {Nature Materials},
  year      = {2021},
  volume    = {20},
  pages     = {1353--1357},
  doi       = {10.1038/s41563-021-01034-y}
}

@article{WilsonOrtiz2024,
  author    = {Wilson, Stephen D. and Ortiz, Brenden R.},
  title     = {\textit{A}\ch{V3Sb5} kagome superconductors},
  journal   = {Nature Reviews Materials},
  year      = {2024},
  volume    = {9},
  number    = {6},
  pages     = {420--432},
  doi       = {10.1038/s41578-024-00677-y}
}

@article{Teng2022,
  author    = {Teng, Xiaokun and Chen, Lebing and Ye, Feng and Rosenberg, Elliott and Liu, Zhaoyu and Yin, Jia-Xin and Jiang, Yu-Xiao and Oh, Ji Seop and Hasan, M. Zahid and Neubauer, Kelly J. and Gao, Bin and Xie, Yaofeng and Hashimoto, Makoto and Lu, Donghui and Jozwiak, Chris and Bostwick, Aaron and Rotenberg, Eli and Birgeneau, Robert J. and Chu, Jiun-Haw and Yi, Ming and Dai, Pengcheng},
  title     = {Discovery of charge density wave in a kagome lattice antiferromagnet},
  journal   = {Nature},
  year      = {2022},
  volume    = {609},
  pages     = {490--495},
  doi       = {10.1038/s41586-022-05034-z}
}

@article{Zhao2021,
  author    = {Zhao, He and Li, Hong and Ortiz, Brenden R. and Teicher, Samuel M. L. and Park, Taka and Ye, Mengxing and Wang, Ziqiang and Wilson, Stephen D. and Zeljkovic, Ilija},
  title     = {Cascade of correlated electron states in the kagome superconductor \ch{CsV3Sb5}},
  journal   = {Nature},
  year      = {2021},
  volume    = {599},
  pages     = {216--221},
  doi       = {10.1038/s41586-021-03946-w}
}

@article{MuSR2022,
  author    = {Mielke, C. and Das, D. and Yin, J.-X. and Liu, H. and Gupta, R. and Jiang, Y.-X. and Medarde, M. and Wu, X. and Lei, H. and Chang, J. and Dai, P. and Si, Q. and Miao, H. and Thomale, R. and Neupert, T. and Shi, Y. and Khasanov, R. and Hasan, M. Z. and Luetkens, H. and Guguchia, Z.},
  title     = {Time-reversal symmetry-breaking charge order in a kagome superconductor},
  journal   = {Nature},
  year      = {2022},
  volume    = {602},
  pages     = {245--250},
  doi       = {10.1038/s41586-021-04327-z}
}

@article{CVSAMRO2021,
  author    = {Xiang, Ying and Li, Qing and Li, Yongkai and Xie, Wei pebbles and Yang, Huan and Wang, Zhiwei and Yao, Yugui and Wen, Hai-Hu},
  title     = {Twofold symmetry of c-axis resistivity in topological kagome superconductor \ch{CsV3Sb5} with in-plane rotating magnetic field},
  journal   = {Nature Communications},
  year      = {2021},
  volume    = {12},
  number    = {1},
  pages     = {6727},
  doi       = {10.1038/s41467-021-27084-z}
}

@article{LiangWu2022,
  author    = {Xu, Yishuai and Ni, Zhuoliang and Liu, Yizhou and Ortiz, Brenden R. and Deng, Qinwen and Wilson, Stephen D. and Yan, Binghai and Balents, Leon and Wu, Liang},
  title     = {Three-state nematicity and magneto-optical Kerr effect in the charge density waves in kagome superconductors},
  journal   = {Nature Physics},
  year      = {2022},
  volume    = {18},
  number    = {12},
  pages     = {1470--1475},
  doi       = {10.1038/s41567-022-01805-7}
}

@article{CVSChiral2022,
  author    = {Guo, Chunyu and Putzke, Carsten and Konyzheva, Svitlana and Huang, Xiang and Gutteck, Matthew and Gao, Bin and Dai, Pengcheng and Gati, Elena and Mohamed, Marisa S. and Zuniga-Cespedes, Nicholas and Sickinger, Christopher and Johnston, Sean and Piatti, Erik and Pasritiya, Rajat and Sassi, Valentina and Shin, Seung-Hoon and Zhang, Qiang and Akhmetov, Alim and Shahi, Purnima and Geibel, Christoph and Moessner, Roderich and Hicks, Clifford W. and Mackenzie, Andrew P. and Moll, Philip J. W.},
  title     = {Switchable chiral transport in charge-ordered kagome metal \ch{CsV3Sb5}},
  journal   = {Nature},
  year      = {2022},
  volume    = {611},
  pages     = {461--466},
  doi       = {10.1038/s41586-022-05127-9}
}

@article{KVSAHE,
author = {Shuo-Ying Yang  and Yaojia Wang  and Brenden R. Ortiz  and Defa Liu  and Jacob Gayles  and Elena Derunova  and Rafael Gonzalez-Hernandez  and Libor Šmejkal  and Yulin Chen  and Stuart S. P. Parkin  and Stephen D. Wilson  and Eric S. Toberer  and Tyrel McQueen  and Mazhar N. Ali },
title = {Giant, unconventional anomalous Hall effect in the metallic frustrated magnet candidate, \ch{KV3Sb5}},
journal = {Science Advances},
volume = {6},
number = {31},
pages = {eabb6003},
year = {2020},
doi = {10.1126/sciadv.abb6003},
URL = {https://www.science.org/doi/abs/10.1126/sciadv.abb6003},
eprint = {https://www.science.org/doi/pdf/10.1126/sciadv.abb6003}}

@article{CVSAHE,
  title = {Concurrence of anomalous Hall effect and charge density wave in a superconducting topological kagome metal},
  author = {Yu, F. H. and Wu, T. and Wang, Z. Y. and Lei, B. and Zhuo, W. Z. and Ying, J. J. and Chen, X. H.},
  journal = {Phys. Rev. B},
  volume = {104},
  issue = {4},
  pages = {L041103},
  numpages = {7},
  year = {2021},
  month = {Jul},
  publisher = {American Physical Society},
  doi = {10.1103/PhysRevB.104.L041103},
  url = {https://link.aps.org/doi/10.1103/PhysRevB.104.L041103}
}

@article{ColossalC-axisCVS,
  title = {Colossal $c$-Axis Response and Lack of Rotational Symmetry Breaking within the Kagome Planes of the \ch{CsV3Sb5} Superconductor},
  author = {Frachet, Mehdi and Wang, Liran and Xia, Wei and Guo, Yanfeng and He, Mingquan and Maraytta, Nour and Heid, Rolf and Haghighirad, Amir-Abbas and Merz, Michael and Meingast, Christoph and Hardy, Fr\'ed\'eric},
  journal = {Phys. Rev. Lett.},
  volume = {132},
  issue = {18},
  pages = {186001},
  numpages = {7},
  year = {2024},
  month = {May},
  publisher = {American Physical Society},
  doi = {10.1103/PhysRevLett.132.186001},
  url = {https://link.aps.org/doi/10.1103/PhysRevLett.132.186001}
}

@article{CVS_nonematicity,
  title = {Absence of ${E}_{2g}$ Nematic Instability and Dominant ${A}_{1g}$ Response in the Kagome Metal \ch{CsV3Sb5}},
  author = {Liu, Zhaoyu and Shi, Yue and Jiang, Qianni and Rosenberg, Elliott W. and DeStefano, Jonathan M. and Liu, Jinjin and Hu, Chaowei and Zhao, Yuzhou and Wang, Zhiwei and Yao, Yugui and Graf, David and Dai, Pengcheng and Yang, Jihui and Xu, Xiaodong and Chu, Jiun-Haw},
  journal = {Phys. Rev. X},
  volume = {14},
  issue = {3},
  pages = {031015},
  numpages = {11},
  year = {2024},
  month = {Jul},
  publisher = {American Physical Society},
  doi = {10.1103/PhysRevX.14.031015},
  url = {https://link.aps.org/doi/10.1103/PhysRevX.14.031015}
}

@article{Saykin2023,
  title = {High Resolution Polar Kerr Effect Studies of \ch{CsV3Sb5}: Tests for Time-Reversal Symmetry Breaking below the Charge-Order Transition},
  author = {Saykin, David R. and Farhang, Camron and Kountz, Erik D. and Chen, Dong and Ortiz, Brenden R. and Shekhar, Chandra and Felser, Claudia and Wilson, Stephen D. and Thomale, Ronny and Xia, Jing and Kapitulnik, Aharon},
  journal = {Phys. Rev. Lett.},
  volume = {131},
  issue = {1},
  pages = {016901},
  numpages = {6},
  year = {2023},
  month = {Jul},
  publisher = {American Physical Society},
  doi = {10.1103/PhysRevLett.131.016901},
  url = {https://link.aps.org/doi/10.1103/PhysRevLett.131.016901}
}

@article{Farhangetal2023,
  author    = {Farhang, Camron and Wang, Jingyuan and Ortiz, Brenden R. and Wilson, Stephen D. and Xia, Jing},
  title     = {Unconventional specular optical rotation in the charge ordered state of Kagome metal $\text{CsV}_3\text{Sb}_5$},
  journal   = {Nature Communications},
  year      = {2023},
  volume    = {14},
  number    = {1},
  pages     = {5326},
  doi       = {10.1038/s41467-023-41080-5},
  publisher = {Springer Science and Business Media LLC}
}

@article{korshunov_softening_2023,
	title = {Softening of a flat phonon mode in the kagome \ch{ScV6Sn6}},
	volume = {14},
	copyright = {2023 The Author(s)},
	issn = {2041-1723},
	url = {https://www.nature.com/articles/s41467-023-42186-6},
	doi = {10.1038/s41467-023-42186-6},
	number = {1},
	urldate = {2026-06-27},
	journal = {Nature Communications},
	author = {Korshunov, A. and Hu, H. and Subires, D. and Jiang, Y. and C{\u a}lug{\u a}ru, D. and Feng, X. and Rajapitamahuni, A. and Yi, C. and Roychowdhury, S. and Vergniory, M. G. and Strempfer, J. and Shekhar, C. and Vescovo, E. and Chernyshov, D. and Said, A. H. and Bosak, A. and Felser, C. and Bernevig, B. Andrei and Blanco-Canosa, S.},
	month = oct,
	year = {2023},
	pages = {6646},
}

@article{cao_competing_2023,
	title = {Competing charge-density wave instabilities in the kagome metal \ch{ScV6Sn6}},
	volume = {14},
	copyright = {2023 The Author(s)},
	issn = {2041-1723},
	url = {https://www.nature.com/articles/s41467-023-43454-1},
	doi = {10.1038/s41467-023-43454-1},
	number = {1},
	urldate = {2026-06-27},
	journal = {Nature Communications},
	author = {Cao, Saizheng and Xu, Chenchao and Fukui, Hiroshi and Manjo, Taishun and Dong, Ying and Shi, Ming and Liu, Yang and Cao, Chao and Song, Yu},
	month = nov,
	year = {2023},
	pages = {7671},
}

@article{tan_abundant_2023,
	title = {Abundant {Lattice} {Instability} in {Kagome} {Metal} \ch{ScV6Sn6}},
	volume = {130},
	url = {https://link.aps.org/doi/10.1103/PhysRevLett.130.266402},
	doi = {10.1103/PhysRevLett.130.266402},
	number = {26},
	urldate = {2026-06-27},
	journal = {Physical Review Letters},
	author = {Tan, Hengxin and Yan, Binghai},
	month = jun,
	year = {2023},
	pages = {266402},
}

@article{ECE_Ikeda,
  title = {{{AC}} Elastocaloric Effect as a Probe for Thermodynamic Signatures of Continuous Phase Transitions},
  author = {Ikeda, M. S. and Straquadine, J. A. W. and Hristov, A. T. and Worasaran, T. and Palmstrom, J. C. and Sorensen, M. and Walmsley, P. and Fisher, I. R.},
  year = {2019},
  month = aug,
  journal = {Review of Scientific Instruments},
  volume = {90},
  number = {8},
  pages = {083902},
  issn = {0034-6748, 1089-7623},
  doi = {10.1063/1.5099924},
  urldate = {2025-05-06},
  langid = {english}
}

@article{twoband_coey,
  title = {Evidence for Two-Band Magnetotransport in Half-Metallic Chromium Dioxide},
  author = {Watts, S. M. and Wirth, S. and Von Moln{\'a}r, S. and Barry, A. and Coey, J. M. D.},
  year = {2000},
  month = apr,
  journal = {Physical Review B},
  volume = {61},
  number = {14},
  pages = {9621--9628},
  issn = {0163-1829, 1095-3795},
  doi = {10.1103/PhysRevB.61.9621},
  urldate = {2025-04-04},
  copyright = {http://link.aps.org/licenses/aps-default-license},
  langid = {english}
}

@article{quantumoscillationsevidence_SVS,
  title = {Quantum Oscillations Evidence for Topological Bands in Kagome Metal \ch{{ScV6Sn6}}},
  author = {Zheng, Guoxin and Zhu, Yuan and Mozaffari, Shirin and Mao, Ning and Chen, Kuan-Wen and Jenkins, Kaila and Zhang, Dechen and Chan, Aaron and Arachchige, Hasitha W Suriya and Madhogaria, Richa P and Cothrine, Matthew and Meier, William R and Zhang, Yang and Mandrus, David and Li, Lu},
  year = {2024},
  month = feb,
  journal = {Journal of Physics: Condensed Matter},
  volume = {36},
  number = {21},
  pages = {215501},
  publisher = {IOP Publishing},
  issn = {0953-8984},
  doi = {10.1088/1361-648X/ad2803},
  urldate = {2025-04-03},
  langid = {english}
}

@article{universalsublinearresistivity_SVS_CVS,
  title = {Universal Sublinear Resistivity in Vanadium Kagome Materials Hosting Charge Density Waves},
  author = {Mozaffari, Shirin and Meier, William R. and Madhogaria, Richa P. and Peshcherenko, Nikolai and Kang, Seoung-Hun and Villanova, John W. and Arachchige, Hasitha W. Suriya and Zheng, Guoxin and Zhu, Yuan and Chen, Kuan-Wen and Jenkins, Kaila and Zhang, Dechen and Chan, Aaron and Li, Lu and Yoon, Mina and Zhang, Yang and Mandrus, David G.},
  year = {2024},
  month = jul,
  journal = {Physical Review B},
  volume = {110},
  number = {3},
  pages = {035135},
  issn = {2469-9950, 2469-9969},
  doi = {10.1103/PhysRevB.110.035135},
  urldate = {2025-04-03},
  langid = {english}
}

@article{SVS_PG,
  title = {Pseudogap Behavior in Charge Density Wave Kagome Material \ch{{ScV6Sn6}} Revealed by Magnetotransport Measurements},
  author = {DeStefano, Jonathan M. and Rosenberg, Elliott and Peek, Olivia and Lee, Yongbin and Liu, Zhaoyu and Jiang, Qianni and Ke, Liqin and Chu, Jiun-Haw},
  year = {2023},
  month = nov,
  journal = {npj Quantum Materials},
  volume = {8},
  number = {1},
  pages = {1--7},
  publisher = {Nature Publishing Group},
  issn = {2397-4648},
  doi = {10.1038/s41535-023-00600-8},
  urldate = {2023-11-08},
  copyright = {2023 The Author(s)},
  langid = {english}
}

@article{tuningSVS_Crdoping,
  title = {Tuning Charge Density Wave of Kagome Metal \ch{{ScV6Sn6}}},
  author = {Yi, Changjiang and Feng, Xiaolong and Kumar, Nitesh and Felser, Claudia and Shekhar, Chandra},
  year = {2024},
  month = may,
  journal = {New Journal of Physics},
  volume = {26},
  number = {5},
  pages = {052001},
  publisher = {IOP Publishing},
  issn = {1367-2630},
  doi = {10.1088/1367-2630/ad4389},
  urldate = {2025-04-02},
  langid = {english}
}

@article{natureofCDW_SVS,
  title = {Nature of Charge Density Wave in Kagome Metal \ch{{ScV6Sn6}}},
  author = {Lee, Seongyong and Won, Choongjae and Kim, Jimin and Yoo, Jonggyu and Park, Sudong and Denlinger, Jonathan and Jozwiak, Chris and Bostwick, Aaron and Rotenberg, Eli and Comin, Riccardo and Kang, Mingu and Park, Jae-Hoon},
  year = {2024},
  month = jan,
  journal = {npj Quantum Materials},
  volume = {9},
  number = {1},
  pages = {1--8},
  publisher = {Nature Publishing Group},
  issn = {2397-4648},
  doi = {10.1038/s41535-024-00620-y},
  urldate = {2025-02-18},
  copyright = {2024 The Author(s)},
  langid = {english}
}

@article{intermediatenematic_SVS,
  title = {Discovery of an Intermediate Nematic State in a Bilayer Kagome Metal \ch{{ScV6Sn6}}},
  author = {Farhang, Camron and Meier, William R. and Lu, Weihang and Li, Jiangxu and Wu, Yudong and Mozaffari, Shirin and Madhogaria, Richa P. and Zhang, Yang and Mandrus, David and Xia, Jing},
  year = 2025,
  month = aug,
  journal = {Nature Communications},
  volume = {16},
  number = {1},
  pages = {7867},
  publisher = {Nature Publishing Group},
  issn = {2041-1723},
  doi = {10.1038/s41467-025-63294-5},
  urldate = {2026-04-20},
  copyright = {2025 The Author(s)},
  langid = {english}
}

@article{CVS_tinypockets,
  title = {Impact of {{Tiny Fermi Pockets}} with {{Extremely High Mobility}} on the {{Hall Anomaly}} in the {{Kagome Metal}} \ch{CsV3Sb5}},
  author = {Liu, S. and Roppongi, M. and Kimata, M. and Ishihara, K. and Grasset, R. and Konczykowski, M. and Ortiz, B. R. and Wilson, S. D. and Yoshimi, K. and Shibauchi, T. and Hashimoto, K.},
  year = 2025,
  month = jul,
  journal = {Physical Review Letters},
  volume = {135},
  number = {5},
  pages = {056502},
  publisher = {American Physical Society},
  doi = {10.1103/d4dw-2v6k},
  urldate = {2026-04-16}
}

@article{svs_discovery,
  title = {Charge {{Density Wave}} in {{Kagome Lattice Intermetallic}} \ch{ScV6Sn6}},
  author = {Arachchige, Hasitha W. Suriya and Meier, William R. and Marshall, Madalynn and Matsuoka, Takahiro and Xue, Rui and McGuire, Michael A. and Hermann, Raphael P. and Cao, Huibo and Mandrus, David},
  year = {2022},
  month = nov,
  journal = {Physical Review Letters},
  volume = {129},
  number = {21},
  pages = {216402},
  publisher = {American Physical Society},
  doi = {10.1103/PhysRevLett.129.216402},
  urldate = {2022-12-24}
}

@article{QuantumOscillationsRevealing_SVS,
  title = {Quantum Oscillations Revealing Topological Band in Kagome Metal \ch{ScV6Sn6}},
  author = {Yi, Changjiang and Feng, Xiaolong and Mao, Ning and Yanda, Premakumar and Roychowdhury, Subhajit and Zhang, Yang and Felser, Claudia and Shekhar, Chandra},
  year = {2024},
  month = jan,
  journal = {Physical Review B},
  volume = {109},
  number = {3},
  pages = {035124},
  issn = {2469-9950, 2469-9969},
  doi = {10.1103/PhysRevB.109.035124},
  urldate = {2024-12-18},
  langid = {english}
}

@article{hasan_svsnematic,
  title = {Van {{Hove}} Annihilation and Nematic Instability on a Kagome Lattice},
  author = {Jiang, Yu-Xiao and Shao, Sen and Xia, Wei and Denner, M. Michael and Ingham, Julian and Hossain, Md Shafayat and Qiu, Qingzheng and Zheng, Xiquan and Chen, Hongyu and Cheng, Zi-Jia and Yang, Xian P. and Kim, Byunghoon and Yin, Jia-Xin and Zhang, Songbo and Litskevich, Maksim and Zhang, Qi and Cochran, Tyler A. and Peng, Yingying and Chang, Guoqing and Guo, Yanfeng and Thomale, Ronny and Neupert, Titus and Hasan, M. Zahid},
  year = 2024,
  month = sep,
  journal = {Nature Materials},
  volume = {23},
  number = {9},
  pages = {1214--1221},
  publisher = {Nature Publishing Group},
  issn = {1476-4660},
  doi = {10.1038/s41563-024-01914-z},
  urldate = {2025-10-29},
  copyright = {2024 The Author(s), under exclusive licence to Springer Nature Limited},
  langid = {english}
}

@article{shapiroMeasurementB1gB2g2016,
  title = {Measurement of the ${{B_{1g}}}$ and ${{B_{2g}}}$ Components of the Elastoresistivity Tensor for Tetragonal Materials via Transverse Resistivity Configurations},
  author = {Shapiro, M. C. and Hristov, A. T. and Palmstrom, J. C. and Chu, Jiun-Haw and Fisher, I. R.},
  year = 2016,
  month = jun,
  journal = {Review of Scientific Instruments},
  volume = {87},
  number = {6},
  pages = {063902},
  issn = {0034-6748},
  doi = {10.1063/1.4953334},
  urldate = {2026-06-09}
}

@misc{svs_strain_Tcdwincrease,
  title = {Symmetry-{{Selective Stabilization}} of {{Charge-Density Wave}} in \ch{ScV6Sn6}},
  author = {Korshunov, A. and Lim, C.-Y. and {Corral-Sertal}, J. and Garbarino, G. and Chernyshov, D. and Rajapitamahuni, A. and Yi, C. and Roychowdhury, S. and Shekhar, C. and Felser, C. and Pardo, V. and Schmidt, Ella M. and {Blanco-Canosa}, S.},
  year = 2026,
  month = jun,
  number = {arXiv:2606.05866},
  eprint = {2606.05866},
  primaryclass = {cond-mat.str-el},
  publisher = {arXiv},
  doi = {10.48550/arXiv.2606.05866},
  urldate = {2026-06-09},
  archiveprefix = {arXiv}
}

@article{modifiedmontgomery,
  title = {Ubiquitous Signatures of Nematic Quantum Criticality in Optimally Doped {{Fe-based}} Superconductors},
  author = {Kuo, Hsueh-Hui and Chu, Jiun-Haw and Palmstrom, Johanna C. and Kivelson, Steven A. and Fisher, Ian R.},
  year = 2016,
  month = may,
  journal = {Science},
  volume = {352},
  number = {6288},
  pages = {958--962},
  publisher = {American Association for the Advancement of Science},
  doi = {10.1126/science.aab0103},
  urldate = {2026-06-09}
}

@article{svs_pressure,
  title = {Destabilization of the {{Charge Density Wave}} and the {{Absence}} of {{Superconductivity}} in \ch{{ScV6Sn6}} under {{High Pressures}} up to 11 {{GPa}}},
  author = {Zhang, Xiaoxiao and Hou, Jun and Xia, Wei and Xu, Zhian and Yang, Pengtao and Wang, Anqi and Liu, Ziyi and Shen, Jie and Zhang, Hua and Dong, Xiaoli and Uwatoko, Yoshiya and Sun, Jianping and Wang, Bosen and Guo, Yanfeng and Cheng, Jinguang},
  year = 2022,
  month = jan,
  journal = {Materials},
  volume = {15},
  number = {20},
  pages = {7372},
  publisher = {Multidisciplinary Digital Publishing Institute},
  issn = {1996-1944},
  doi = {10.3390/ma15207372},
  urldate = {2022-10-26},
  copyright = {http://creativecommons.org/licenses/by/3.0/},
  langid = {english}
}

@article{huPhononPromotedCharge2024,
  title = {Phonon Promoted Charge Density Wave in Topological Kagome Metal \ch{{ScV6Sn6}}},
  author = {Hu, Yong and Ma, Junzhang and Li, Yinxiang and Jiang, Yuxiao and Gawryluk, Dariusz Jakub and Hu, Tianchen and Teyssier, J{\'e}r{\'e}mie and Multian, Volodymyr and Yin, Zhouyi and Xu, Shuxiang and Shin, Soohyeon and Plokhikh, Igor and Han, Xinloong and Plumb, Nicholas C. and Liu, Yang and Yin, Jia-Xin and Guguchia, Zurab and Zhao, Yue and Schnyder, Andreas P. and Wu, Xianxin and Pomjakushina, Ekaterina and Hasan, M. Zahid and Wang, Nanlin and Shi, Ming},
  year = 2024,
  month = feb,
  journal = {Nature Communications},
  volume = {15},
  number = {1},
  pages = {1658},
  publisher = {Nature Publishing Group},
  issn = {2041-1723},
  doi = {10.1038/s41467-024-45859-y},
  urldate = {2026-06-13},
  copyright = {2024 The Author(s)},
  langid = {english}
}

@article{cvs_fermisurfacenesting,
  title = {Twofold van {{Hove}} Singularity and Origin of Charge Order in Topological Kagome Superconductor \ch{{CsV3Sb5}}},
  author = {Kang, Mingu and Fang, Shiang and Kim, Jeong-Kyu and Ortiz, Brenden R. and Ryu, Sae Hee and Kim, Jimin and Yoo, Jonggyu and Sangiovanni, Giorgio and Di Sante, Domenico and Park, Byeong-Gyu and Jozwiak, Chris and Bostwick, Aaron and Rotenberg, Eli and Kaxiras, Efthimios and Wilson, Stephen D. and Park, Jae-Hoon and Comin, Riccardo},
  year = 2022,
  month = mar,
  journal = {Nature Physics},
  volume = {18},
  number = {3},
  pages = {301--308},
  publisher = {Nature Publishing Group},
  issn = {1745-2481},
  doi = {10.1038/s41567-021-01451-5},
  urldate = {2026-04-13},
  copyright = {2022 The Author(s), under exclusive licence to Springer Nature Limited},
  langid = {english}
}

@article{hiddenmagnetism_svs,
	title = {Hidden magnetism uncovered in a charge ordered bilayer kagome material \ch{ScV6Sn6}},
	volume = {14},
	copyright = {2023 The Author(s)},
	issn = {2041-1723},
	url = {https://www.nature.com/articles/s41467-023-43503-9},
	doi = {10.1038/s41467-023-43503-9},
	number = {1},
	urldate = {2026-06-27},
	journal = {Nature Communications},
	publisher = {Nature Publishing Group},
	author = {Guguchia, Z. and Gawryluk, D. J. and Shin, S. and Hao, Z. and Mielke III, C. and Das, D. and Plokhikh, I. and Liborio, L. and Shenton, J. Kane and Hu, Y. and Sazgari, V. and Medarde, M. and Deng, H. and Cai, Y. and Chen, C. and Jiang, Y. and Amato, A. and Shi, M. and Hasan, M. Z. and Yin, J.-X. and Khasanov, R. and Pomjakushina, E. and Luetkens, H.},
	month = nov,
	year = {2023},
	pages = {7796},
}

@article{kapitulnik_SVS,
  title = {High-Resolution Polar {{Kerr}} Effect Studies of \ch{CsV3Sb5} and \ch{ScV6Sn6} below the Charge Order Transition},
  author = {Saykin, David R. and Jiang, Qianni and Liu, Zhaoyu and Shekhar, Chandra and Felser, Claudia and Chu, Jiun-Haw and Kapitulnik, Aharon},
  year = 2026,
  month = jun,
  journal = {Physical Review B},
  volume = {113},
  number = {24},
  pages = {L241104},
  publisher = {American Physical Society},
  doi = {10.1103/xfv5-3z9t},
  urldate = {2026-06-18}
}

@article{rattlingchain,
  title = {Tiny {{Sc Allows}} the {{Chains}} to {{Rattle}}: {{Impact}} of {{Lu}} and {{Y Doping}} on the {{Charge-Density Wave}} in \ch{{ScV6Sn6}}},
  shorttitle = {Tiny {{Sc Allows}} the {{Chains}} to {{Rattle}}},
  author = {Meier, William R. and Madhogaria, Richa Pokharel and Mozaffari, Shirin and Marshall, Madalynn and Graf, David E. and McGuire, Michael A. and Arachchige, Hasitha W. Suriya and Allen, Caleb L. and Driver, Jeremy and Cao, Huibo and Mandrus, David},
  year = 2023,
  month = sep,
  journal = {Journal of the American Chemical Society},
  volume = {145},
  number = {38},
  pages = {20943--20950},
  publisher = {American Chemical Society},
  issn = {0002-7863},
  doi = {10.1021/jacs.3c06394},
  urldate = {2026-06-19}
}

\end{document}